\documentclass{appolb}
\usepackage{graphicx}
\usepackage{amsmath}
\usepackage{amssymb}
\usepackage[usenames]{color}
\usepackage[colorlinks=true,linkcolor=blue,citecolor=blue,urlcolor=blue]{hyperref}
\usepackage{cite}

\begin{document}
\title{Non-perturbative features of the three-gluon vertex in Landau gauge\thanks{Presented by M. Vujinovic at the Workshop
``Excited QCD 2014'', Bjela\v snica Mountain, Sarajevo, Bosnia-Herzegovina, February 2-8, 2014}}

\author{Milan Vujinovic, Reinhard Alkofer
\address{Institut f\"{u}r Physik, Karl-Franzens-Universit\"{a}t Graz, Universit\"{a}tsplatz 5, \\8010 Graz, Austria} \\ [3mm]
Gernot Eichmann, Richard Williams
\address{Institut f\"{u}r Theoretische Physik, Justus-Liebig-Universit\"{a}t Giessen, \\35392 Giessen, Germany}
}
\maketitle
\begin{abstract}
We present a calculation of the three-gluon vertex from its Dyson-Schwinger equation in Landau-gauge Yang-Mills theory.
All tensor structures are considered and back-coupled self-consistently.
Within the chosen truncation, two-loop diagrams as well as diagrams containing Green's functions beyond the
primitively divergent ones are neglected.
Only the three-gluon vertex is chosen to be dynamical; the other propagators and vertex functions are provided as separate
solutions of their Dyson-Schwinger equations or by Ansatz.
For both scaling- and decoupling-type solutions we observe, in agreement with other studies, a sign
change in the tree-level tensor dressing at a non-perturbative scale.
\end{abstract}

\section{Introduction}

In Yang-Mills theories the three- and four-gluon vertices are those two primitively-divergent Green functions through which the gluon
self-inter\-action of non-Abelian QCD is inherent.
Thus, it is expected that many interesting non-perturbative phenomena are encoded within.
Further, they are expected to play a prominent role in phenomenological studies of QCD and QCD-like theories, especially in the context of bound states.
In these proceedings we focus upon the three-gluon vertex as discussed in Ref.~\cite{Eichmann:2014xya}.

There have been numerous quantitative and qualitative investigations of this vertex.
General tensor decompositions consistent with the Slavnov-Taylor identities of Yang-Mills theory were considered
in \cite{Kim:1979ep, Ball:1980ax}.
There are extensive studies within perturbation theory~\cite{Davydychev:1996ws, Binger:2006sj}, but also
non-perturbative calculations of infrared critical exponents for scaling \cite{Alkofer:2004it, Alkofer:2008dt} and
decoupling~\cite{Aguilar:2013vaa} solutions.
Together with lattice results for a particular projection of the vertex~\cite{Maas:2007uv, Cucchieri:2008qm}, 
this knowledge is sufficient for the construction of
vertex models used in Dyson-Schwinger equation (DSE) based calculations of the gluon propagator \cite{Huber:2012kd}.
The fact that the gluon propagator obtained in this way is, in turn, in good agreement with the respective
lattice results adds to the evidence that carefully constructed models can capture the essential features of vertex functions.

However, there are possibly important properties of the three-gluon vertex which can only be probed through explicit calculation.
For example, the aforementioned lattice results for the vertex indicate that its leading structure
becomes  negative at some infrared (IR) scale.
On the lattice, however, this behaviour is clearly seen only in two and three dimensions whereas
the evidence in four dimensions is not yet compelling.
On the other hand,
a zero crossing of the tree level dressing has been convincingly identified and analysed in four-dimensional continuum studies
\cite{Pelaez:2013cpa,Blum:2014gna,Eichmann:2014xya}.
A sign change might play an important role when the three-gluon vertex is used for studies of hadronic observables. Possible investigations in this direction
include studies of mesons beyond rainbow-ladder~\cite{Fischer:2009jm,Williams:2014iea}, excited states, glueballs and hybrids.

The three-gluon vertex is also relevant for other strongly coupled gauge theories, for example technicolor~\cite{Weinberg:1979bn}.
In the context of modern technicolor theories, it is of particular importance to learn about the behaviour of fundamental Green's functions in the case of (nearly)-conformal dynamics~\cite{Holdom:1981rm, Appelquist:1986an}. Thus far, this influence has only been assessed for propagators
\cite{Maas:2011jf, August:2013jia, Alkofer:2014taa}. Gaining knowledge on the behaviour of vertex functions near or inside
the conformal window is expected to improve our understanding of strongly coupled gauge theories in general.
In this regard the quark-gluon vertex is very interesting, as the components of this vertex which break chiral symmetry are expected to vanish in an exactly conformal theory. However, through its DSE the three-gluon vertex is a potential driving term for the quark-gluon vertex.

\section{Three-Gluon-Vertex Dyson-Schwinger equation}

\begin{figure}[t]

\centerline{%
\includegraphics[width=1.00\textwidth]{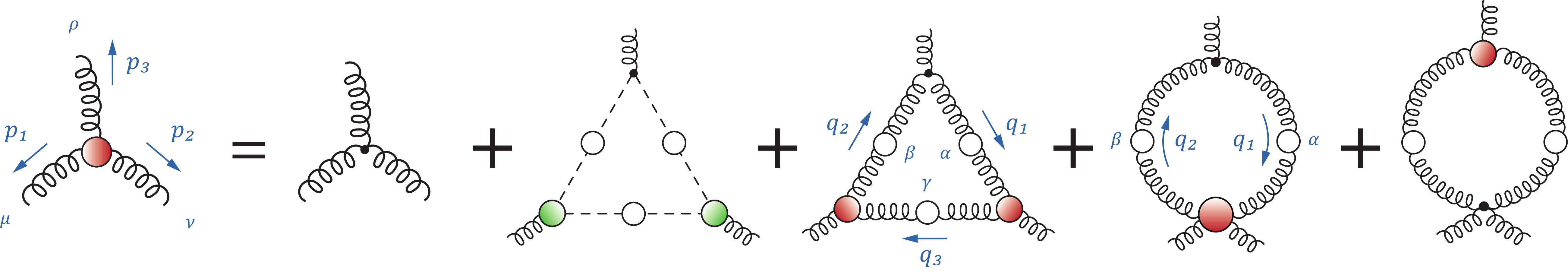}}
\caption{Truncated DSE for three-gluon vertex. Dashed lines represent ghosts and wiggly ones gluons.}
\label{Fig:F2H}
\end{figure}

The three-gluon vertex DSE is Bose symmetric. This is not clearly mani\-fest upon truncation, but can be restored by
considering all permutations with respect to the interchange of external gluon legs. The truncated DSE, wherein non-primitively
divergent vertices and two-loop terms are neglected, is given in Fig.~\ref{Fig:F2H}.

\subsection{Ghost/gluon input and vertex models}

In Landau gauge the ghost and gluon propagators are of the form
\begin{equation}
  D_G(k) = -\frac{G(k^2)}{k^2}, ~~~~ D^{\mu\nu}(k) = T^{\mu\nu}_k\frac{Z(k^2)}{k^2}\,,
\end{equation}
where
$G(k^2)$ and $Z(k^2)$ are, respectively, the ghost and gluon dressing functions and
$T^{\mu\nu}_k = \delta^{\mu\nu} - k^\mu k^\nu/k^2$ is the transverse projector.
Owing to transversality of the gluon propagator in Landau gauge, only the transverse components of the
three-gluon vertex need to be retained.

In Ref.~\cite{Blum:2014gna} the coupled system of DSEs for ghost and gluon propagators as well as the
three-gluon vertex was solved.
It was found that the back-reaction of the three-gluon vertex onto the ghost/gluon system does not change
the propagators appreciably.
In that study only the tree-level structure of the three-gluon vertex was considered. Here we perform a
complimentary investigation: we keep the ghost and gluon propagators fixed, but retain all tensor structures of the
three-gluon vertex.
As detailed below, we quantitatively confirm the dominance of the tree-level tensor component but  also find some interesting features in the sub-leading terms.

In a first step we construct fit functions for $G(k^2)$ and $Z(k^2)$ based on the calculation of \cite{Fischer:2002hna}.
(NB: These fit functions are discussed in detail in  \cite{Eichmann:2014xya}.
In \cite{Fischer:2002hna}  only scaling solutions were studied;
here we also consider the decoupling solutions within the same truncation scheme.)
The ghost-gluon vertex is substituted by its bare form, $\Gamma^{\mu}_{gh} = \Gamma^{\mu}_{gh,0}$,
which is justified from many studies in Landau gauge~\cite{Cucchieri:2004sq,Schleifenbaum:2004id,Cucchieri:2008qm, Huber:2012kd}.
For the four-gluon vertex a model is used which employs only the tree-level tensor structure: $\Gamma_{4g}^{\mu\nu\rho\sigma}(p_1,p_2,p_3,p_4) =
f_{4g}(x)~\Gamma_{4g,0}^{\mu\nu\rho\sigma}$, with $x \sim p_1^2 + p_2^2 + p_3^2 + p_4^2$.
The function $f_{4g}$ matches the known behaviour of the tree-level dressing
in the ultraviolet \cite{Celmaster:1979km, Pascual:1980yu} and IR regions \cite{Alkofer:2004it, Kellermann:2008iw}. We allow further freedom in the modelling by shifting the dressing
function by a constant: $f_{4g}(x) + (0\ldots0.6)$.

\begin{figure}[t]
\centerline{%
\includegraphics[width=13cm]{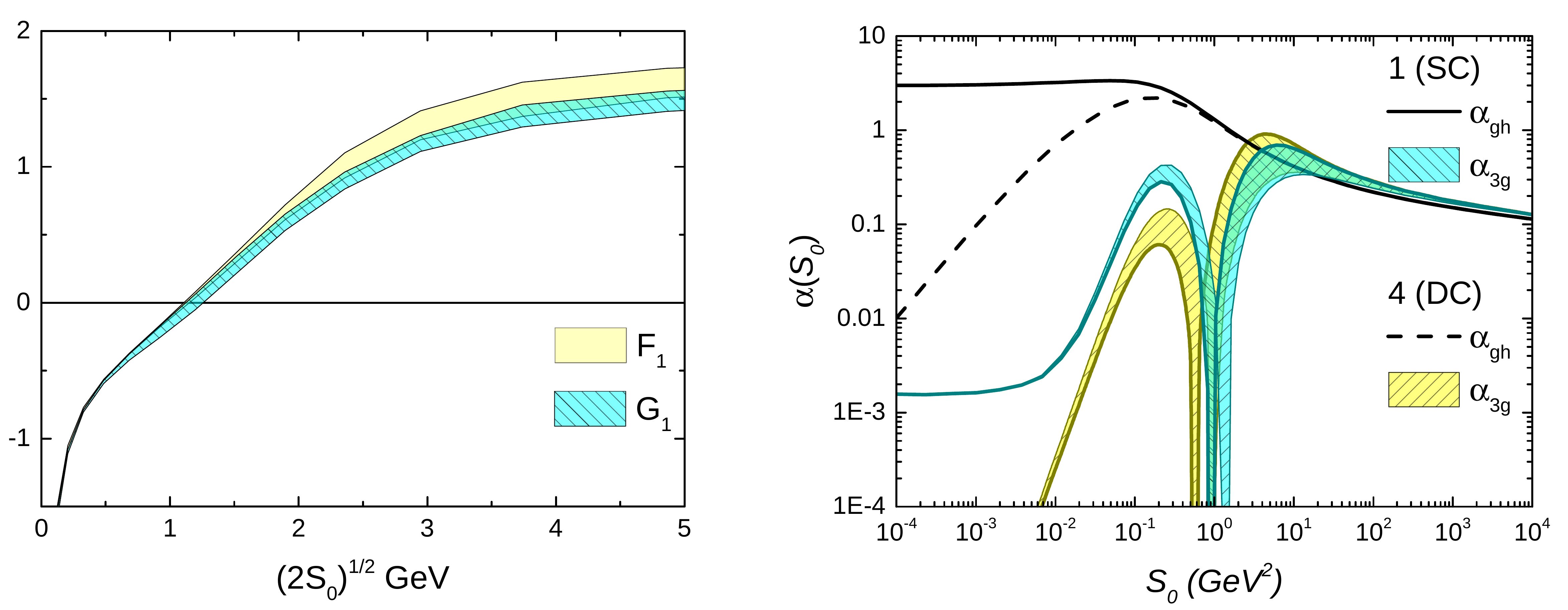}}
\caption{\textit{Left panel}: Tree-level dressing $F_1$ and the corresponding projection $G_1$ as used in lattice studies. \textit{Right panel}: Running couplings of Eq. (3)
for scaling (SC) and decoupling (DC) scenarios. The bands represent a variation in the four-gluon vertex model.}
\label{Fig:F5H}
\end{figure}

\section{Numerical results}

The chosen conventions for the vertex kinematics are shown in Fig.~\ref{Fig:F2H}.
In general one needs three independent variables to describe the three-gluon vertex.
In these proceedings we will only show results for the
symmetric limit $p_1^2 = p_2^2 = p_3^2$ and use correspondingly
the Bose-symmetric combination of momenta, $\mathcal{S}_0 = (p_1^2 + p_2^2 + p_3^2)/6$.

\subsection{Tree-level dressing functions and running couplings}

 In the left panel of Fig.~\ref{Fig:F5H}
  we show the tree-level dressing $F_1$ together with the function $G_1$ which is the projection that has been calculated on the lattice:
\begin{equation}
G_1(p_1, p_2, p_3) = \frac{\Gamma_{3g,0}^{\mu\nu\rho}(p_1, p_2, p_3)T^{\mu\mu'}_{p_1} T^{\nu\nu'}_{p_2}T^{\rho\rho'}_{p_3}\Gamma_{3g}^{\mu'\nu'\rho'}(p_1, p_2, p_3)}
{\Gamma_{3g,0}^{\mu\nu\rho}(p_1, p_2, p_3)T^{\mu\mu'}_{p_1}T^{\nu\nu'}_{p_2}T^{\rho\rho'}_{p_3}\Gamma_{3g,0}^{\mu'\nu'\rho'}(p_1, p_2, p_3)}\, .
\end{equation}
The functions $F_1$ and $G_1$ look similar and show a sign change at the same scale $p \approx 1.1\ldots 1.4$ GeV.
 Deviations between $F_1$ and $G_1$ originate from an admixture of sub-leading components within $G_1$, which contribute at the 10 $\%$ level \cite{Eichmann:2014xya}.
However, the location of the zero crossing is considerably higher than what has been seen in lattice studies \cite{Maas:2007uv, Cucchieri:2008qm}.
The dependence of the zero-crossing location on the four-gluon vertex model
 is rather mild, see Fig.~\ref{Fig:F5H}. However, there is still the possibility of
corrections coming from neglected two-loop terms in the vertex DSE.

In the right panel of Fig.~\ref{Fig:F5H} we show the results for the ghost-gluon and three-gluon running couplings which are given by
\begin{align}
  &\alpha_{gh} = \alpha(\mu^2)Z(p^2)G^2(p^2)\,, \nonumber \\
  &\alpha_{3g} = \alpha(\mu^2)Z^3(p^2)F^2_1(p^2)\,,
\end{align}
with $\alpha(\mu^2) = g^2(\mu^2)/4\pi$. The couplings $\alpha_{3g}$ and $\alpha_{gh}$ agree well in the UV
(the slight mismatch comes from the neglect of two-loop terms in the vertex DSE), but they look quite differently in the IR:
there is no unique
non-perturbative definition of the running coupling. In the scaling scenario both couplings have an infrared fixed point, in agreement with the exact IR
analysis \cite{Alkofer:2004it}.

\section{Conclusions and outlook}

We performed an investigation of the three-gluon vertex in Landau gauge Yang-Mills theory. All relevant tensor
structures were back-coupled self-consistently. We demonstrated within our truncation that the 
tree-level vertex dressing function has the expected
infrared behaviour in both scaling and decoupling scenarios.
It exhibits a zero crossing whose location is higher than one would expect from (lower-dimensional) lattice studies.
While its location is only mildly affected by the details of the four-gluon vertex, missing two-loop diagrams might still play a larger role.

An open question is therefore the true location of this zero.
Thus, future studies will focus upon resolving these details,
along with the impact that this peculiar feature of the three-gluon vertex has
on hadron phenomenology as well as Green's functions in the conformal window
of strongly interacting theories.

\bigskip

\textbf{Acknowledgments.} We acknowledge support by the German Science Foundation (DFG) under project number DFG TR-16,
the Austrian Science Fund (FWF) under project numbers M1333-N16, J3039-N16 and P25121-N27, and from the Doktoratskolleg ``Hadrons in Vacuum, Nuclei and Stars'' (FWF) DK W1203-N16.

\end{document}